\documentclass[11pt,twoside]{article}
\usepackage{asp2014}

\aspSuppressVolSlug
\resetcounters

\begin{document}

\title{Radio data archives round table}

% full name: Katharina Lutz
\author{K.~A.~Lutz$^1$, J. Dempsey$^2$, Y.~G. Grange$^3$, M. Kettenis$^4$, M. Lacy$^5$, and C. Schollar$^6$}
\affil{$^1$CDS, Observatoire Astronomique de Strasbourg,  Universit\'{e} de Strasbourg, CNRS - UMR 7550, Strasbourg, France; \email{research@katha-lutz.de}}
\affil{$^2$CSIRO Information Management and Technology, GPO Box 1700 Canberra, ACT 2601, Australia}
\affil{$^3$ASTRON, the Netherlands Institute for Radio Astronomy, Oude Hoogeveensedijk 4, 7991 PD Dwingeloo, The Netherlands}
\affil{$^4$JIVE, Dwingeloo, Netherlands}
\affil{$^5$NRAO, Charlottesville, VA, USA}
\affil{$^6$SKA SA, Cape Town, State/Province, South Africa}
% remove/add as you need

% remove/add authors as you need
\paperauthor{K.~A.~Lutz}{research@katha-lutz.de}{0000-0002-6616-7554}{CDS, CNRS, Observatoire Astronomique de Strasbourg}{}{Strasbourg}{}{67000}{France}
\paperauthor{J.~Dempsey}{james.dempsey@csiro.au}{0000-0002-4899-4169}{CSIRO}{Information Management and Technology}{Canberra}{ACT}{2601}{Australia}
\paperauthor{Y.~G.~Grange}{grange@astron.nl}{0000-0001-5125-9539}{ASTRON}{}{Dwingeloo}{}{7991PD}{the Netherlands}
\paperauthor{M. Kettenis}{}{}{JIVE}{}{Dwingeloo}{}{}{Netherlands}
\paperauthor{M. Lacy}{}{}{NRAO}{}{Charlottesville}{}{}{USA}
\paperauthor{C. Schollar}{}{}{SKA SA}{}{Cape Town}{}{}{South Africa}
% remove/add as you need

% leave these next few aindex lines commented for the editors to enable them. Use Aindex.py to generate them for yourself.
% first presenting author should be the first entry for bold-facing the author index page-reference
%\aindex{Lutz,~K.~A.}
% \aindex{Dempsey, J.}
% \aindex{Grange, Y.~G.}
% \aindex{Kettenis, M.}
% \aindex{Lacy, M.}
% \aindex{Schollar, C.}
% remove/add as you need

% leave the ssindex lines commented for the editors to enable them, use Index.py to suggest yours
%\ssindex{FOOBAR!conference!ADASS 2020}
%\ssindex{FOOBAR!organisations!ASP}
%\ssindex{astronomy!radio}
%\ssindex{astronomy!radio!VLBI}
%\ssindex{astronomy!radio!interferometry}
%\ssindex{astronomy!radio!single-dish}
%\ssindex{archives}

% leave the ooindex lines commented for the editors to enable them, use ascl.py to suggest yours
%\ooindex{FOOBAR, ascl:1101.010}
% \ooindex{CASA, ascl:1107.013}

\begin{abstract}

With SKA precursor and pathfinder operations in full swing, radio and (sub-)mm astronomy is entering the era of super big data. The big questions is how to make (sub-)mm and radio data available to the astronomical community, preferably using FAIR (findable, accessible, interoperable and re-useable; \citealp{2016_Wilkinson_FAIR}) principles. There are already a lot of efforts going on around the globe: facilities such as ALMA, LOFAR, MWA, NRAO and ASKAP are already publishing much of their imaging data in the form of "science ready" products, SKA regional centres are being formed and a radio astronomy interest group has been initiated within the IVOA. In addition, also non-imaging data, like timing, pulsar or beam forming data, needs to be available to the community as well. This BoF intended to bring everyone interested in this topic around one virtual table to hear about and discuss the following questions: What is the status of efforts to expose both visibility and science ready data? What is already there, maybe has been used for decades by traditional observatories? What is  still missing? Where do we want to go next?
\end{abstract}

\section{Introduction}

We organised this birds-of-a-feather (BoF) session to open a forum for the (radio) astronomy community to discuss the current status and future of radio data archives.
%Due to the virtual format of the conference, the discussion took place both on a zoom webinar and in the conference discord channels. In addition, conference participants were also able to follow the discussion in a youtube stream, on this platform, however, there was no possibility to contribute to the discussion.
The BoF started with presentations from the panel members. These presentations included  reports on current status of a variety of archives, visions for the future and open questions. After the presentations the floor was opened to other radio data providers and users to comment and highlight their own projects (Sec.\ref{sec:current}). Throughout the open BoF discussion open questions and important topics became clear (Sec.~\ref{sec:open}).

\section{Current state of Radio Data archives around the world}
\label{sec:current}
\textbf{Presentation: radio data archive and the Virtual Observatory} \\
The recently founded Radio Astronomy interest group \footnote{https://wiki.ivoa.net/twiki/bin/view/IVOA/IvoaRadio} within the International Virtual Observatory Alliance (IVOA) is now ramping up its activities. The aim of this interest group is to discuss how radio data can be served through the protocols developed by the IVOA. One use case, based on which the protocols might be further developed, could be the stacking of atomic hydrogen emission in the SKA era: get a catalogue of galaxies with measured redshifts, move on to the SKA archive, download cubes around the redshift and spatial location of each galaxy, and stack the profiles. If then we were interested in the average molecular gas content of these galaxies, we should be able to send an almost identical query to ALMA. With Virtual Observatory (VO) protocols implemented, these kind of queries should be easy to execute for astronomers, archive and software agnostic and scalable to larger samples of sources.

\textbf{Presentation: what types of data to publish and how to describe data types}\\
There is a multitude of data, which can be published in archives (raw, calibrated visibilities, diagnostic plots, science ready data) but there is also quite some data that is not shared, e.\,g. because these were obtained in non-standard observation modes. Ideally archives would serve all these data and also support PIs of small observing projects to publish their reduced and generated data. However, there are limiting factors like limited storage space and people power.

In the future, we should try to publish visibilities as long as it is feasible. Furthermore, archives should be connected to science platforms, which is facilitated if using VO protocols. In addition, for science ready and processed data like legacy surveys, a description of the data provenance is essential. Within JIVE, they aim to provide Jupyter notebooks with the data. With these notebooks, users should be able to retrace the reduction process and tweak it where desired. This would already provide science platform functionalities. For HTTP data transfers and the use of VO protocols, it is furthermore important to standardise the media types (MIME types) of data stored in (radio) archives. These media types are available for FITS and UVFITS but not for e.\,g. Measurement Sets, FITS-IDI, PSR-FITS, etc.. This would for example allow science analysis platforms to suggest appropriate tools for processing the data.

\textbf{Presentation: the CSIRO radio data archive and what to serve}\\
Within CSIRO, three archives are serving radio data: CASDA, which focuses on science ready data from ASKAP, PSRDA, which is the Parkes pulsar time series data archive, and ATOA, which serves raw visibilities from ATCA, Parkes and Mopra. In discussing, which data to serve, the CSIRO radio data archives realised that not only storage space but also the capacity of astronomers to interact with large datasets are limiting factors. For ASKAP within CASDA the decision has been made to only store calibrated but no raw visibilities, which is controversial as the calibration pipeline has to be extremely reliable and trustworthy. In the future, no spectral line visibilities will be stored anymore. For continuum observations, storage of \textit{averaged}, calibrated visibilities is envisioned in the long term. In the future, observations with the Parkes CryoPAF will produce 80TB of data per day, of which likely only discoveries can be stored. Thinking about the discovery of fast radio burst (see also Sec.~\ref{sec:open}), this approach might miss data which is not interesting at the time of observation but highly interesting at a later point in time.

\textbf{Presentation: the Astron radio data archive}\\
The majority of data stored in the Astron archive comes from LOFAR. LOFAR data and Apertif calibrated visibilities are stored on tape, which makes a staging process inevitable before they can be access for analysis. Even after staging the sheer data size makes it challenging to bring the data to the user. (Near) Science ready data from Apertif are available through a dedicated web interface and also through VO services as of the day of the BoF.

\textbf{Presentation: the MeerKAT radio data archive}\\
MeerKAT aims at storing and serving visibility data for the foreseeable future. To do so, the archive system consists of two units: the medium term, spinning disc archive and the long term tape archive. The medium term archive is a Ceph (S3 like) object and can thus be especially well access through http/s. To be able to host all data, a project to expand the archive to 30PB of spinning disc and 36PB of tape storage is ongoing.

In addition also the MeerKAT visibility format has been optimised to work with S3 like infrastructure and is thus easy to load and inspect via the network. A dedicated python library for data access and tools to convert from the MeerKAT visibility format to better known data formats are made available.

\textbf{Open discussion: Other archives}\\
Both ALMA and MWA are serving data through dedicated web interfaces and VO services. ALMA will primarily provide science ready data and serve both raw and calibrated visibilities only as auxiliary data. For now all data can still be stored. MWA focuses on raw visibilities and voltages. Their storage system is not big enough, so they set up a first-in first-out queue but people can reserve data to be kept a little longer. The CADC hosts several radio/mm collections in the same data model that supports to the VO data models and access protocols. The collections are co-located and integrated in the CANFAR science platform.

\section{Open questions and issues}
\label{sec:open}
\textbf{Backward compatibility} \\
Radio data archives contain observations obtained over the last decades. However, with the development of new software packages and file format standards, some of these data might need "old" versions of software packages or even operating systems to be processed. \citet{P11-220} demonstrated how the East Asian ALMA Regional Centre is overcomes this issue with a containerised reduction pipeline. After the BoF itself more information and expertise regarding containerised reduction pipelines has been exchanged on the discord channel.

\textbf{How to make visibilities more useful and discoverable}\\
Most archives plan to serve visibilities as much as possible. Unlike for other observations, the creation of images or cubes from visibilities can yield different results depending on the processing parameters.
In addition, the detection of fast radio burst archival raw radio data, and the distribution of calibrated visibilities instead of data cubes to the Event Horizon Telescope science teams make important usecases for serving visibilities in archives.
%In the case of the Event Horizon Telescope observations of the black hole in the centre of M\,87, calibrated visibilities were provided to different science teams, who then processed these according to their primary science goal. Furthermore, fast radio burst were discovered in archival and reprocessed radio data.
% While it is true that fast radio burst were discovered in archival data, that wasn't visibility data but voltage data. It can be seen as another example of "raw" data that may not always be archived.
The best metadata description of visibilities is, however, still under investigation. Due to the sheer amount of data, visibilities of one observation/experiment are almost always stored in multiple files. The metadata of one of these files are not able to provide information on the sensitivity of the entire observation. Furthermore, there is nothing like thumbnails available to judge the amount of artefacts. In addition, the spatial resolution, the field of view and largest observable angular scale are dependent on the imaging process. In summary, concepts to describe the metadata of visibilities in a way that allows to query for them in a meaningful way are still under investigation.

Another route to simplifying the access to visibilities is to offer science platforms. On these platforms, astronomers would be able to reprocess the data on the server, where the data are stored. To make it even easier for users, one could imagine to provide calibrated visibilities to process, and/or a simple menu-driven process.

\textbf{Quality measures}\\
When publishing science ready data or calibrated visibilities, archives need to ensure a certain level of quality. This not only simplifies the usage of these data by astronomers, who's main expertise is not in radio astronomy. These measures will also increase the trust of experienced radio astronomers in science ready products especially in the case that (raw) visibilities are not available any more. Observatories shared their approaches: NRAO have defined a quality standard, which datasets pass or not, and provide information on necessary steps to calibrate a dataset. Within ASKAP, the science teams were asked to provide quality tests and measures against which data are checked. ALMA have automated quality control within their reduction pipeline and experiment with machine learning support. The ASTRON archive team works on quantifying the quality of the uv-coverage, so that this parameter might be queried on.

\textbf{File formats}\\
%Again due to the sheer data volume of visibility data, there are efforts to store visibility data in a different data format than would be used for data analysis. This is also important to optimise the (speed of) data analysis e.\,g. through parallel computing. For example, the ALMA archive uses ASDM to store visibilities, which are then analysed in CASA Measurement Sets. The MeerKAT archive have developed their own visibility format, which is optimised for their Amazon S3-like access to the data storage. They are currently working on combining the new format and approaches with what is known and trusted by the community.
To fully exploit data archives, common data formats are essential. For visibility data the Measurement Set (MS) has established itself as a de-facto data model. Its underlying Casacore Table Data System (CTDS) data format is not fully standardised though. Due to the sheer data volume of visibility data, there are efforts to store visibility data in a different data formats, both for archival purposes and for analysis. This is important to optimise the (speed of) data analysis e.\,g. through parallel computing. For example, the ALMA archive uses ASDM to store visibilities, which are then analysed in CASA MS. The MeerKAT archive have developed their own visibility format, which is optimised for their Amazon S3-like access to the data storage. To optimise the analysis, NRAO and SARAO are developing data access layers based on Dask and Xarray.

\textbf{Publish science ready data of small surveys}\\
While observatories are currently mostly focusing on serving science ready data products from large surveys, there was also a point made that science ready data products from small surveys should be made available. In many cases these data are only presented in Figures of scientific publications. However, data cubes or images, which can be reused by the astronomical community, would be highly valuable. Options to publish these data are the VizieR associated data service or (considered) initiatives by the observatories.
%One option to publish these data is provided by the VizieR associated data service. Also observatories are looking into ways to make this kind of services available for reduced data observed with their telescopes.

In \textbf{Conclusion} we had a very lively discussion about many topics. A workshop, where different data providers could develop things together would certainly be a great idea for the future.
\bibliography{B9-50}

% \bookpartphoto[width=1.0\textwidth]{foobar.eps}{FooBar Photo (Photo: Any Photographer)}

\end{document}